\documentclass[prl,aps,twocolumn,amssymb,showpacs]{revtex4}
\usepackage{amsmath,epsfig}
\usepackage{graphicx}

\begin{document}

\title{Extrinsic noise driven phenotype switching in a self-regulating gene}
\author{Michael Assaf$^{1,}$\footnotemark[1]}
\author{Elijah Roberts$^{2,}$\footnotemark[1]}
\author{Zaida Luthey-Schulten$^{3,4}$}
\author{Nigel Goldenfeld$^4$}
\affiliation{$^1$Racah Institute of Physics, Hebrew University of Jerusalem, Jerusalem 91904, Israel}
\affiliation{$^2$Department of Biophysics, Johns Hopkins University, Baltimore, MD 21218} %
\affiliation{Department of Chemistry$^3$ and Physics$^4$, University of Illinois at Urbana-Champaign, Urbana, Illinois 61801}
\begin{abstract}
Due to inherent noise in intracellular networks cellular decisions can be random, so genetically identical cells can display different phenotypic behavior even in identical environments. Most previous work in understanding the decision-making process has focused on the role of intrinsic noise in these systems. Yet, especially in the high copy-number regime, extrinsic noise has been shown to be much more significant. Here, using a prototypical example of a bistable self-regulating gene model, we develop a theoretical framework describing the combined effect of intrinsic and extrinsic noise on the dynamics of stochastic genetic switches. Employing our theory and Monte Carlo simulations, we show that extrinsic noise not only significantly alters the lifetimes of the phenotypic states, but can induce bistability in unexpected regions of parameter space, and may fundamentally change the escape mechanism. These results have implications for interpreting experimentally observed heterogeneity in cellular populations and for stochastic modeling of cellular decision processes.
\end{abstract}
\maketitle

{\it Introduction.$\;$} Noise-driven switching between coexisting metastable states plays a key role in many systems in physics,  chemistry, and biology~\cite{Horsthemke1984,Hanggi,Gardiner2004hsm,CollinsIdoReview}. Besides thermal or intrinsic noise (IN) that drives switching~\cite{GeneSwitchingTheory}, such systems often experience extrinsic or environmental noise (EN) from the noisy environment or from being coupled to another fluctuating system~\cite{Horsthemke1984}. Noise-driven escape from a metastable state while under the influence of EN has been previously studied in the context of population biology and population genetics (see \textit{e.g.}~\cite{Karlin1974tfs,Leigh1981alp,Kamenev2008hce}). Here, \textit{e.g.}, it has been shown that delta-correlated as well as colored EN can drastically decrease the population's mean extinction time~\cite{Leigh1981alp,Kamenev2008hce}. Moreover, recently there has been a large effort to predict the onset of EN-driven critical transitions and regime shifts in ecosystems, see \textit{e.g.}~\cite{Scheffer}.

In cellular biology, most studies of gene expression dynamics, including our own treatments~\cite{ourintrinsic}, have focused on the role of IN (reviewed in~\cite{intrinsic}). Recently, however, gene expression under EN has also come under study~\cite{extrinsic,Taniguchi2010qec}, where EN has been experimentally confirmed to be one of the dominant sources of variation in protein copy number, particularly above copy numbers of ${\cal O}(10)$~\cite{Taniguchi2010qec}. In studies of genetic switches, EN has been shown to induce bistability~\cite{Samoilov2005,Shahrezaei2008cef}, vary the distribution tails~\cite{Shahrezaei2008cef} and modify switching times~\cite{Kessler2011}. Yet, previous studies have not provided fundamental insight as to the interplay between IN and EN in the switching process, {\itshape i.e.}, how the mean switching times (MSTs) and switching paths deviate according to EN strength, correlation time and statistics. Elucidating the relationship between IN and EN during switching is crucial to understanding how EN affects population heterogeneity in bistable systems, which is of importance when studying, {\itshape e.g.}, bet-hedging strategies like bacterial persistence~\cite{Balaban2004bpp}.

In this Letter we study the contributions of IN and EN to noise-driven switching in a simple self-regulating genetic circuit with positive feedback.
Employing a semi-classical theory we perform a systematic study of the effect of EN statistics, magnitude and correlation time, on the switch's stochastic dynamics. In particular, we derive expressions for the MSTs as functions of the EN strength and correlation time, and also study how EN can induce bistability in an otherwise monostable system. All analytical results are corroborated by extensive Monte-Carlo (MC) simulations. Our main conclusion is that EN correlation time plays a significant role in determining both the stability of the metastable state and the mechanism of escape. This strongly indicates that in biological systems, where the correlation time is thought to be long, phenotype switching may be driven primarily by EN.

\textit{Model.$\;$}
Our  analysis relies on the model of a self-regulating gene (SRG), with positive feedback due to the production rate depending on the state. Let $n(t)$ be the protein copy number and $N$ be the protein abundance in the $hi$ state. Proteins are produced at a rate $f(n)$, which is any Hill-like function, and decay with rate $1$. The mean protein concentration $\bar{x}(t)=\bar{n}(t)/N$ satisfies
\begin{equation}\label{MF}
\dot{\bar{x}}=f(\bar{x})-\bar{x}.
\end{equation}
For simplicity we take $f(x)=\alpha_0+(1-\alpha_0)\theta(x-x_0)$, where $\theta(x)$ is the Heaviside step function, and $\alpha_0<x_0<1$. Eq.~(\ref{MF}) leads to a bistable system with three fixed points $x_1<x_2<x_3$, where $x_1=\alpha_0$ and $x_3=1$ are attracting fixed points of the $low$ and $hi$ states respectively, while $x_2=x_0$ is repelling. Typically, $\alpha_0\ll 1$ so $x_3\gg x_1$.

To account for IN, we employ the master equation for $P_n(t)$ - the probability to find $n$ proteins at time $t$:
\begin{equation}\label{master}
\dot{P}_n=f(n-1)P_{n-1}+(n+1)P_{n+1}-[f(n)+n]P_n.
\end{equation}
For simplicity we focus on the weak-noise regime $1-x_0\ll 1$, where (without loss of generality) the ``switching barrier" between the $hi$ and $low$ states is small. In this regime Eq.~(\ref{master}) is accurately approximated~\cite{Assaf2010ems} by the following  Fokker-Planck equation (FPE) for the probability $P(x,t)$ to find  concentration $x$ at time $t$~\cite{Gardiner2004hsm}:
\begin{equation}\label{FPE}
\partial_t P=-\partial_x\{[f(x)-x]P\} + 1/(2N)\,\partial^2_x\{[f(x)+x]P\}.
\end{equation}
Starting from the vicinity of the $hi$ state, the system rapidly forms a quasi-stationary distribution (QSD) about the $hi$ state, which slowly leaks through the unstable point $x=x_0$~\cite{Assaf2010ems,Dykman1994lfo,Assaf2006stm}. In general, the metastable state decays as $P(x,t)\simeq \pi(x)e^{-t/\tau}$ where $\pi(x)$ is the QSD and $\tau$ is the MST. Employing the WKB ansatz $\pi(x)\sim e^{-NS(x)}$ for the QSD, where $S(x)$ is called the action and $p(x)\equiv S'(x)$ is called the momentum~\cite{Dykman1994lfo}, Eq.~(\ref{FPE}) gives rise to a stationary Hamilton-Jacobi equation (HJE) $H(x,p_x)=0$ with Hamiltonian
\begin{equation}\label{HAM}
H(x,p_x)=p_x[f(x)-x]+(p_x^2/2)[f(x)+x].
\end{equation}
Switching occurs along the  zero-energy trajectory $p_x(x)=-2[f(x)-x]/[f(x)+x]$ of (\ref{HAM}). For $x_0 < x\leq 1$, $p_x(x)=-2(1-x)/(1+x)$, which for $1-x_0\ll 1$ satisfies $|p_x(x)|\ll 1$. This yields $S(x)=\int^x p_x(x')dx'=2[x-2\ln(1+x)]$, and the QSD around $x=1$: $\pi(x)\sim e^{-N[S(x)-s(1)]}$ with standard deviation $\sigma_{in}=N^{-1/2}$. Therefore, since $\tau_{hi\to low}\sim \pi(x_0)^{-1}$~\cite{Assaf2006stm,Assaf2010ems}, we have~\cite{MST}
\begin{equation}\label{MSTnonoise}
\ln \tau_{hi\to low}\!\simeq\! N[S(x_0)\!-\!S(1)]\!\simeq\! (N/2)(1\!-\!x_0)^2\!\equiv\! \Delta S_0.
\end{equation}
which is applicable as long as $\sigma_{in}=N^{-1/2}\ll 1-x_0$.

Next, we incorporate EN in the form of one or more fluctuating parameters. We assume that cell-to-cell variability in transcription and translation rates causes the protein production rate to fluctuate. In the $hi$ state the production rate then becomes $\alpha_1(t)=1+\xi(t)$, where $\xi(t)$ is fluctuating with finite correlation time. As we are interested in the $hi\to low$ transition  we ignore fluctuations in $\alpha_0\ll 1$. We take $\xi(t)$ to be Ornstein-Uhlenbeck (OU) noise \cite{Gardiner2004hsm}: positively correlated Gaussian noise with zero mean, variance $\sigma_{ex}^2$ and correlation time $\tau_c$, satisfying  $\langle \xi(t)\xi(t')\rangle=\sigma_{ex}^2 e^{-|t-t'|/\tau_c}$. The OU process satisfies the following Langevin equation
\begin{equation}\label{lannoise}
\dot{\xi}=-\xi/\tau_c+\sqrt{2\sigma_{ex}^2/\tau_c}\;\eta(t),
\end{equation}
where $\eta$ is white Gaussian noise, $\langle \eta(t)\eta(t')\rangle=\delta(t\!-\!t')$~\cite{variance}. Here, $\sigma_{ex}^2$ and $\tau_c$ are characteristic of the environment and the cell's regulatory network and are generally unknown. Non-Gaussian statistics for EN have also been proposed \cite{Shahrezaei2008cef}, but further theoretical and experimental work is needed to uncover the source and form of EN.

\begin{figure}
\includegraphics{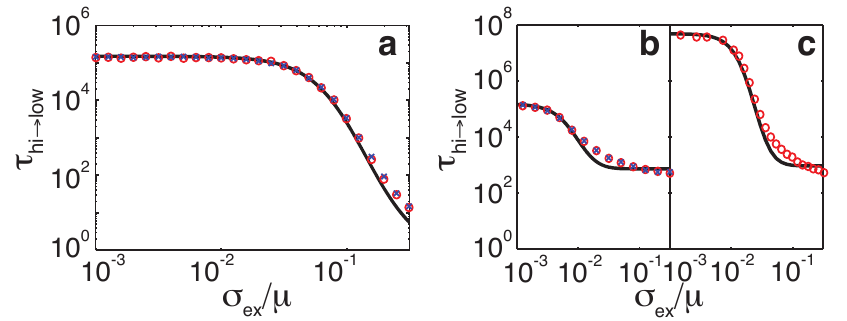}
\caption{(Color online) $\tau_{hi\to low}$ versus the relative strength for white $\tau_c=10^{-2}$ (a) and long-correlated $\tau_c=10^3$ (b+c) EN. MC simulations (the symbols) for noise in the birth (o) and death ($\times$) terms (that are indistinguishable), are compared with theory (lines): Eq.~(\ref{linnoise}) (a) and Eq.~(\ref{corrmstlongnoise}) (b+c). Here $N=5000$,  $\alpha_0=0.01$, and $x_0=0.93$ (a+b) or $x_0=0.915$ (c).}
 \label{fig1}
\end{figure}

To study the interplay between IN and EN, we combine Eq.~(\ref{lannoise}) with the underlying IN dynamics [Eq.~(\ref{FPE})]. Defining the fluctuating production rate $\tilde{f}(x,\xi)=\alpha_0+(1-\alpha_0+\xi)\theta(x-x_0)$, drift term $A(x,\xi)=\tilde{f}(x,\xi)-x$, diffusion coefficient $B(x,\xi)=\tilde{f}(x,\xi)+x$, and the EN and IN variance ratio, $V\equiv \sigma_{ex}^2/\sigma_{in}^2=N\sigma_{ex}^2$, we obtain a 2-D FPE for the \textit{joint probability} $P(x,\xi,t)$ to find concentration $x$ and noise magnitude $\xi$ at time $t$~\cite{extinction,1-2species}:
\begin{eqnarray}\label{2DFPE}
&&\hspace{-4mm}\partial_x P=-\partial_x\left\{A(x,\xi)P\right\}+\partial_{\xi}\left\{(\xi/\tau_c)P\right\}\nonumber\\
&&\hspace{-5mm}+1/(2N)\,\partial^2_x\left\{B(x,\xi)P\right\}+1/(2N)\,\partial^2_{\xi}\left\{(2V/\tau_c)P\right\}.
\end{eqnarray}

Employing the WKB ansatz $P(x,\xi)\!\sim\! e^{-NS(x,\xi)}$ for the QSD, Eq.~(\ref{2DFPE}) yields a HJE: $H(x,\xi,p_x,p_{\xi})=p_x A(x,\xi)-\xi p_\xi/\tau_c+(p_x^2/2)B(x,\xi)+p_\xi^2V/\tau_c=0$, with momenta $p_x\equiv\partial_x S$ and $p_\xi\equiv\partial_\xi S$.
The HJE can be solved by considering the Hamilton equations $\dot{x}_i=\partial_{p_i}H$ and $\dot{p}_i=-\partial_{x_i}H$:
\begin{eqnarray}\label{hameqs}
&&\hspace{-4mm}\dot{x}=A+p_x B\;,\;\;\;\dot{p}_x=-p_x[\partial_x A+(p_x/2)\partial_x B]\nonumber\\
&&\hspace{-4mm}\ddot{\xi}\simeq \xi/\tau_c^2-2p_xV/\tau_c,
\end{eqnarray}
where we have combined the equations for $\dot{\xi}$ and $\dot{p}_{_{\xi}}$ to a single equation for $\ddot{\xi}$ and kept terms up to ${\cal O}(p_x)\ll 1$.

Eqs.~(\ref{hameqs}) can be solved numerically for generic noise, which yields the corresponding action function $S(x,\xi)=\int p_x(x,\xi)dx+p_{_{\xi}}(x,\xi)d\xi$, and QSD. Analytical progress can be made in two limits: short-correlated white noise $\tau_c\ll 1$, and long-correlated adiabatic noise $\tau_c\gg 1$.

\begin{figure}
\includegraphics{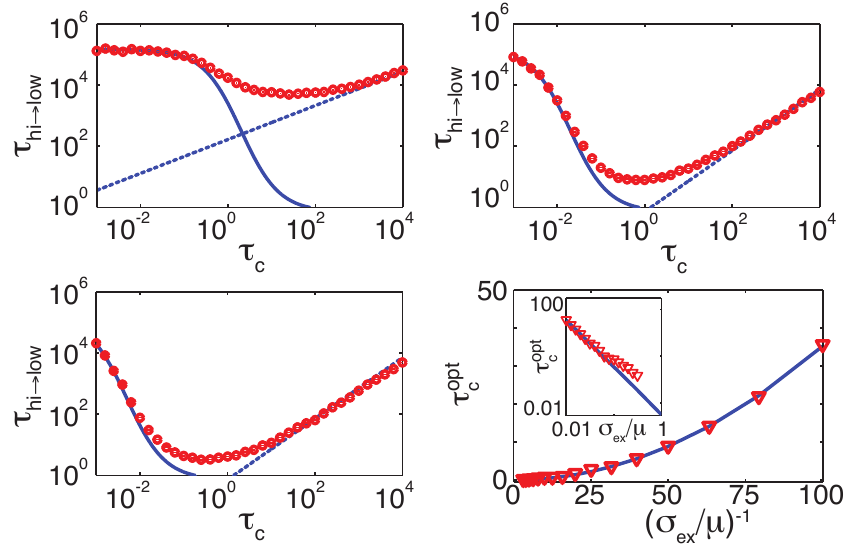}
\caption{(Color online) $\tau_{hi\to low}$ as function of $\tau_c$ for various EN strengths: $\sigma_{ex}/\mu=0.01$ (top left), $\sigma_{ex}/\mu=0.1$ (top right), and $\sigma_{ex}/\mu=0.2$ (bottom left). The lines are the analytical predictions of Eq.~(\ref{linnoise}) (solid) and  Eq.~(\ref{corrmstlongnoise}) (dotted). The lower right panel shows $\tau_c^{opt}$ versus the relative EN strength. Numerical results (symbols)  agree well with Eq.~(\ref{tauopt}) (line) multiplied by $0.72$. Inset shows the clear scaling of $\tau_c^{opt}\sim (\sigma_{ex}/\mu)^2$.}
 \label{fig2}
\end{figure}

For white EN, we neglect $\ddot\xi$ in the third of Eqs.~(\ref{hameqs})~\cite{Kamenev2008hce}, which yields $\xi\simeq 2p_xV\tau_c$. Substituting $\xi$ into the first of Eqs.~(\ref{hameqs}), we find for $x>x_0$: $\dot{x}=f(x)-x+2p_xV\tau_c+p_x[f(x)+x+2p_xV\tau_c]$, which originates from an \textit{effective white-noise Hamiltonian}: $H\simeq p_x[f(x)-x]+(p_x^2/2)[f(x)+x+2V\tau_c]$, where we have neglected ${\cal O}(p_x^3)$ terms. Solving $H=0$, we find $p_x(x)=-2(1-x)/(1+x+2V\tau_c)$, which yields the MST in the white-EN regime
\begin{equation}\label{linnoise}
\ln \tau_{hi\to low}\simeq \Delta S_0/(1+V \tau_c).
\end{equation}
Eq.~(\ref{linnoise}) is confirmed by MC simulations~\cite{simmethods}, see Figs. \ref{fig1}+\ref{fig2}. In Fig.~\ref{fig2} and below, $\mu$ denotes the QSD's average.

Now, to deal with long-correlated EN, we note that when $\tau_c\gg 1$, during the rare fluctuation that takes the system from the $hi$ to the $low$ state, the system samples an almost constant value of the noise $\xi=\xi_0$~\cite{Kamenev2008hce}. For a constant $\xi_0$, the $hi$ fixed point becomes $1+\xi_0$. The optimal value of $\xi_0$ is found by minimizing the cost of switching given noise realization $\xi_0$,
$\ln \tau_{hi\to low}(\xi_0)\simeq  (N/2)(1-x_0+\xi_0)^2$, against the (absolute value of the) statistical weight of $\xi$, $N\xi_0^2/(2V)$. By doing so, we find
$\xi^{opt}=-(1-x_0)V/(1+V)$, where $|\xi^{opt}|<1-x_0$ as expected.
Plugging $\xi^{opt}$ into $\tau_{hi\to low}(\xi_0)$ we find~\footnote{This result can be equivalently obtained by integrating over $\tau_{hi\to low}^{-1}(\xi_0)\sim e^{-(N/2)(1-x_0+\xi_0)^2}$, with the Gaussian weight $e^{-N\xi_0^2/(2V)}$ of $\xi_0$. Using the saddle-point approximation, we recover $\xi_{opt}$, and consequently, Eq.~(\ref{mstlongnoise}).}
\begin{equation}\label{mstlongnoise}
\ln \tau_{hi\to low}\simeq \Delta S_0(1+V)^{-2}.
\end{equation}
For strong EN, $V\gg 1$, Eq.~(\ref{mstlongnoise}) holds when $\Delta S_0 \gg V^{2}$, which can only be satisfied when $\sigma_{ex}\ll 1-x_0$.

What happens when $\sigma_{ex}\gtrsim 1-x_0\gg \sigma_{in}$? Here, IN can be neglected, and the MST turns out to be dominated solely by EN. Namely, the MST can be approximated by the mean first passage time $T(x)$ it takes the OU process to reach position $x$ starting from $x=0$ at $t=0$. Using Eq.~(\ref{lannoise}), $T(x)$ is governed by the following equation~\cite{Gardiner2004hsm}:
\begin{equation}
(\sigma_{ex}^2/\tau_c)T''(x)-(x/\tau_c)T'(x)=-1,
\end{equation}
with boundary conditions $T(0)=0$ and $T'(\infty)=0$, whose solution is $T(x)=\tau_c\varphi(x,\sigma_{ex})$. Here, $\varphi(x,\sigma_{ex})=(\pi/2) \mathrm{Erfi}(z)-z^2\,_2\!F_2\left[\{1,1\},\left\{\frac{3}{2},2\right\},z^2\right],$
where $z= x/(\sqrt{2}\sigma_{ex})$, $\mathrm{Erfi}(z)=2/\sqrt{\pi}\!\int_0^{z}e^{y^2}dy$, and $_2\!F_2(\{\},\{\},x)$ is the generalized hypergeometric function. The MST is obtained by plugging $x=1-x_0$: $\tau_{hi\to low}\simeq T(1-x_0)\simeq \gamma\tau_c$, with $\gamma={\cal O}(1)$ for $\sigma_{ex} \simeq 1-x_0$.

This analysis gives rise to a correction in Eq.~(\ref{mstlongnoise}) for the MST in the adiabatic regime $\tau_c\gg 1$. Since, $\ln\tau_{hi\to low} \simeq \ln \tau_c$ at $\sigma_{ex}\gtrsim 1-x_0$, and $\ln\tau_{hi\to low} = \Delta S_0$ at $\sigma_{ex}=0$, by defining $\lambda=\ln\tau_c$, Eq.~(\ref{mstlongnoise}) becomes
\begin{equation}\label{corrmstlongnoise}
\ln \tau_{hi\to low}\simeq \lambda+(\Delta S_0-\lambda)(1+V)^{-2}.
\end{equation}
Eq.~(\ref{corrmstlongnoise}) compares well with MC numerics, see Figs.~\ref{fig1}+\ref{fig2}.

As can be seen in Fig.~\ref{fig2}, for given EN strength $\sigma_{ex}$ there exists an optimal EN correlation time $\tau_c$ for which the MST is minimal. In order to calculate $\tau_c^{opt}$ we add the white- and adiabatic-noise contributions [Eqs.~(\ref{linnoise}) and (\ref{corrmstlongnoise})] for the MST, and differentiate the result with respect to $\tau_c$. For $1-x_0\ll 1$,
we find
\begin{equation}\label{tauopt}
\tau_c^{opt}\simeq (1-x_0)^2/\sigma_{ex}^2,
\end{equation}
whose dependence on $\sigma_{ex}$ is confirmed by Fig.~\ref{fig2}.
\\
\textit{Noise in the degradation rate.$\;$}
We now consider the case where the degradation rate is fluctuating as $1+\xi(t)$. Here, the corresponding FPE is given by Eq.~(\ref{2DFPE}) with $A(x,\xi)=f(x)-x(1+\xi)$ and $B(x,\xi)=f(x)+x(1+\xi)$.

In the white EN regime the optimal path for switching at $x>x_0$ becomes $p_x(x)=-2(1-x)/(1+x+2x^2V\tau_c)$. This yields a MST that coincides with Eq.~(\ref{linnoise}), see Fig.~\ref{fig1}, since the $x^2$ factor in the denominator of $p_x(x)$ approximately equals $1$ along the integration regime $1-x_0\ll 1$. For adiabatic EN, the $hi$ fixed point becomes $x_3=(1+\xi_0)^{-1}$. This again yields after some algebra
$\xi^{opt}=(1-x_0)V/(1+V)$, which coincides up to a minus sign with $\xi^{opt}$ when the production rate is fluctuating. Therefore, we recover Eq.~(\ref{mstlongnoise}), see Fig.~\ref{fig1}.
\\
\textit{Noise-induced bistability.$\;$}
\begin{figure}
\includegraphics{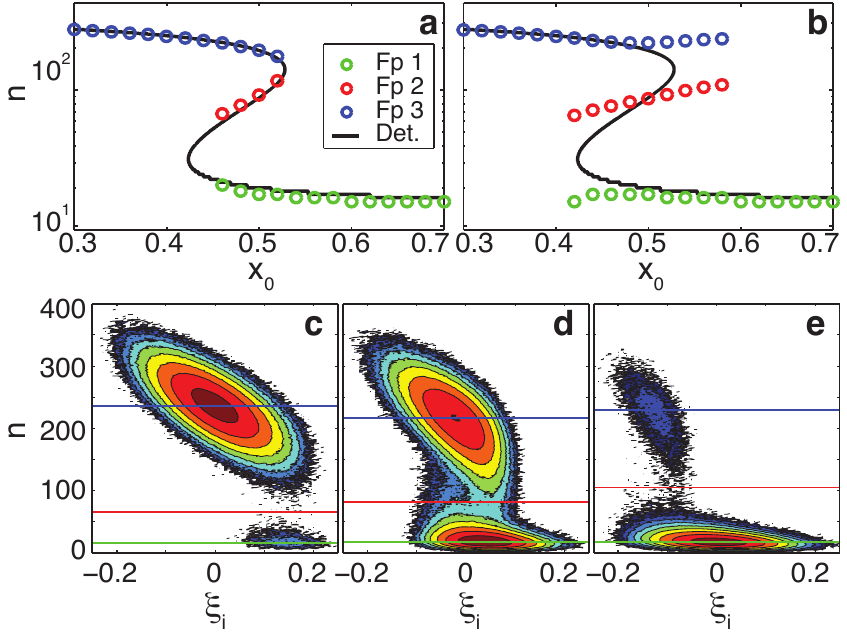}
\caption{(Color online) (a+b) The location of the stochastic fixed points of Eq.~(\ref{MF}) with $f(x)$ given by~(\ref{prodrate}), for EN with  $\tau_c=10^3$ and $\sigma_{ex}/\mu=0.01$ (left) and $\sigma_{ex}/\mu=0.05$ (right). The solid line shows the deterministic fixed points. (c-e) The steady-state 2-D PDFs of finding protein number $n$ and instantaneous EN magnitude $\xi_{i}$ for $\sigma_{ex}/\mu=0.05$ and $x_0=0.42$, $0.48$, and $0.56$, respectively. Here $N=300$, and $\alpha_0=0.05$.}
 \label{fig3}
\end{figure}
To study how EN affects bistability in the SRG model we worked with a modified production rate $f(x)$ that allowed bidirectional transitions between the $low$ and $hi$ states with MSTs that were reachable using MC simulations. Instead of a step-function for $f(x)$ in Eq.~(\ref{MF}), we took
\begin{equation}\label{prodrate}
f(x)=\alpha_0+(1-\alpha_0)x^2/(x^2+x_0^2),
\end{equation}
where $\alpha_0\ll 1$. Given $\alpha_0$ this system is bistable over a range of $x_0$ values. We are interested in how the range of bistability varies under the influence of EN, and also how the latter affects the MSTs and the steady state probability distribution functions (PDFs). To answer these questions we ran MC simulations with a degradation rate $1+\xi(t)$ in the adiabatic limit with $\tau_c=10^3$.

To determine the effect of the EN on the bistability range, we calculated the PDFs at various $x_0$ values from long-time simulations and extracted the position(s) of the sole maximum (monostable) or the two maxima separated by a minimum (bistable). These values we interpreted as stochastic equivalents to the deterministic fixed points. For very weak EN, the stochastic and deterministic fixed points generally agree with only small deviations. Yet, as the EN strength increases the locations of the stochastic fixed points undergo a dramatic departure from their deterministic locations. The example shown in the upper panels of Fig.~\ref{fig3} demonstrates that even for a modest EN strength of $\sigma_{ex}/\mu=0.05$, the range of $x_0$ over which the system is bistable has greatly increased. This effect becomes more pronounced as the EN strength further increases. For $\sigma_{ex}/\mu=0.2$ the system was bistable over the entire range of $x_0$ sampled (0.3-0.7).

To further investigate the change in switching behavior, we calculated the 2-D PDFs of finding protein number $n$ and instantaneous fluctuation magnitude $\xi_i$. The lower panels of Fig.~\ref{fig3} show that, for strong EN, $\xi_i$ has a direct impact on the state of the system. Fig.~\ref{fig3}(d) shows a case where the system is deterministically bistable. Here, when $\xi_i$ is relatively weak the system undergoes noise-driven switching as expected. However, when the degradation reaction is sampling the highest rates, the system exists only in the low state and vice versa. When the EN drives the degradation rate to one of its extremes the system switches deterministically to the appropriate stable state. This effect appears in the 2-D PDFs as two alternate switching paths: when $\xi_i > 0$, there is a $hi\to low$ pathway for leakage of probability, but when $\xi_i < 0$ there is a separate $low\to hi$ leakage path. Thus, the system's bistability is not only a consequence of stochastic switching between states, but also of EN driving the system between different regions of parameter space with alternate fixed point configurations.

Fig.~\ref{fig3}(c+e) show the case where the system is deterministically monostable. When $\xi_i$ is low one can see that the system behaves as though it has a single fixed point. However, when a large fluctuation occurs in the correct direction it can shift the system into a region of parameter space that is bistable; the fluctuations induce bistability in the system. This effect gives rise to the greatly increased bistability range observed in the simulations.

\begin{figure}
\includegraphics{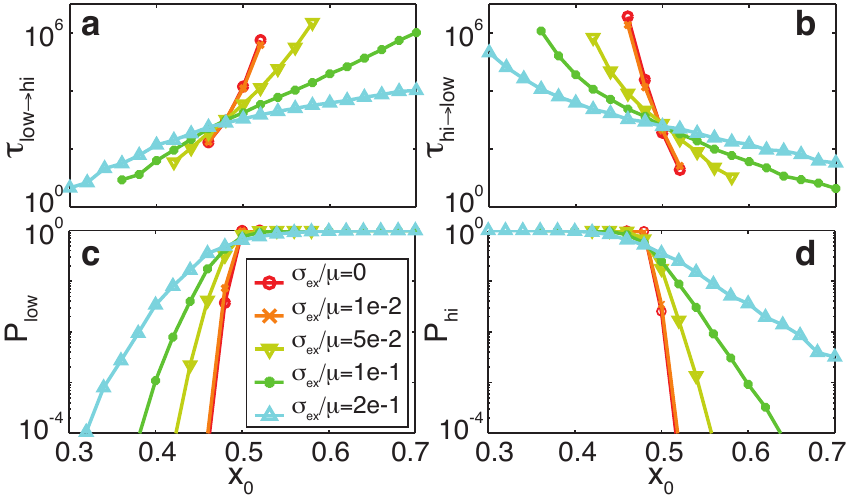}
\caption{(Color online) (a) $\tau_{low\to hi}$ and (b) $\tau_{hi\to low}$ for various EN strengths and $\tau_c=10^3$. The population fraction $P_{low}$ (c) and $P_{hi}$ (d) in the low and high states in steady state.}
 \label{fig4}
\end{figure}

Finally, we calculated the MSTs for different EN strengths. Fig.~\ref{fig4} upper panels show that as the EN magnitude increases, the steepness of the curve as a function of $x_0$ is reduced for both $\tau_{low\to hi}$ and $\tau_{hi\to low}$. Such changes in the MSTs serve to make the less favorable state more populated across a wide range of $x_0$ values. To illustrate this effect, the lower panels of Fig.~\ref{fig4} show the probability of the system being in the $low$ or $hi$ state, calculated as $P_{low}=\tau_{hi\to low}^{-1}/(\tau_{hi\to low}^{-1}+\tau_{low\to hi}^{-1})$ and  $P_{hi}=1-P_{low}$, respectively. Here one can see that as the EN magnitude is increased, not only does the absolute range of bistability expand but so does the range at which the population is macroscopically heterogeneous ({\itshape e.g.} 1 part in 100). The tails of these probabilities decrease much more slowly than a system with only IN.
\\
\textit{Conclusions.$\;$}
Considered in the context of a population of cells, our analysis of a simple SRG model shows that EN is one of the primary drivers of phenotype switching. Switching times can be lowered by multiple orders of magnitude and the mechanism of switching may not be strictly IN-driven, as previously assumed. If we interpret $x_0$ in our model as an environmental input ({\itshape e.g.}, the concentration of an inducer or antibiotic), then the parameter range at which a cellular population will exhibit macroscopic levels of heterogeneity is greatly expanded by EN. Also, by showing how EN can modify the tails of bistable PDFs, our theory provides an interpretation for experimental observations of cells persisting in lowly populated phenotypes across unexpected conditions.



Although this study was based on a simple SRG model, the general results apply to more complex genetic switches where EN is present in many kinetic rates. Stochastic models of cellular decision making will need to account for EN if they are to correctly recover switching times and trajectories. However, the major roadblock is the lack of experimental data regarding the properties of EN. It may be possible to use our theory to deconvolute the effects of IN and EN on switching from switching trajectories of individual cells subject to external fluctuations. We plan to explore such possibilities in the future.

\textit{Acknowledgements.$\;$} We acknowledge support from the NSF via the CPLC at UIUC (PHY-0822613) and from the DOE Office of Science (BER) (E.~R. and Z.~L.-S.) under contract number DE-FG02-10ER6510.
\\
\noindent{$^*$M.~A. and E.~R. contributed equally to this work.}


\begin{thebibliography}{10}
\bibitem{variance} In  Langevin equation~(\ref{lannoise}) $\eta(t)$ can be defined as the $dt\to 0$ limit of the temporally uncorrelated normal random variable with mean $0$ and variance $1/dt$.
\bibitem{1-2species} Here we have increased the problem's dimensions from 1 to 2 by considering the noise magnitude as an \textit{effective additional species} and coupling it to the protein species.
\bibitem{simmethods} Monte Carlo simulations with extrinsic noise were performed using a modified version of the next-reaction variant of the Gillespie algorithm~\cite{GillespieGibsonBruck}. Briefly, a pseudo-reaction fires at intervals much less than the extrinsic noise correlation time and at each firing the $\xi$ term in the propensity of the fluctuating reaction is updated~\cite{GillespieOU} as if it had been following the OU process satisying Eq.~(\ref{lannoise}).
\end{thebibliography}
\end{document}